\documentclass[twoside,leqno,twocolumn]{article}

\usepackage{ltexpprt}

\newcommand{\R}{{\sf R\hspace*{-1.561ex}\rule{0.14ex}{1.6ex}\hspace*{1.561ex}}}
\newcommand{\N}{{\sf N\hspace*{-1.67ex}\rule{0.14ex}{1.6ex}\hspace*{1.67ex}}}
\newcommand{\Np}{{\N_{\scriptscriptstyle +}}}
\newcommand{\Rp}{{\R_{\scriptscriptstyle +}}}

\newcommand{\qed}{%
  \vspace{.06em}\noindent\fbox{\rule{0em}{.01em}\rule{.01em}{0em}}\vspace{0.5em}}

\newenvironment{linearprogram}[2]{
\samepage

\begin{eqnarray} \nonumber
& & \mbox{#1\ } #2 \\ \nonumber
& & \mbox{subject to\ }\left\{ \nonumber
\begin{array}{rcl@{\hspace{0.2in}}l}}{
\end{array}\right.
\end{eqnarray}
}

\makeatletter
\newcommand{\mathfn}[1]{{\operator@font #1}}
\makeatother

\newtheorem{guarantee}{Guarantee}[section]

\newcommand{\mymathfn}[1]{\mathfn{\mathrm{#1}}}

\newcommand{\cost}{\mymathfn{\tt cost}} 
\newcommand{\dist}{\mymathfn{\tt dist}} 
\newcommand{\chern}{\mymathfn{\tt ch}} 
\newcommand{\ce}{\Phi}          
\newcommand{\pe}{\widehat{\ce}} 

\newcommand{\giv}{\,|\,} 

\newcommand{\inp}[1]{{\bf input:} {\em #1}\\} 
\newcommand{\out}[1]{{\bf output:} {\em #1}\vspace{0.1in}\\} 

\newcommand{\tab}{~~~~}

\newcommand{\rnd}[1]{\tilde{#1}} 

\newcommand{\dtm}[1]{\hat{#1}}  
\newcommand{\opt}[1]{{#1}^{*}}  

\newcommand{\comment}[1]{\hfill{\small\em ... #1}} 

\newcommand{\figskip}{\vspace*{1.1ex}}

\begin{document}

\title{\Large $K$-Medians, Facility Location, and the Chernoff-Wald Bound}
\author{Neal E. Young\thanks{%
    Research partially funded by NSF CAREER award CCR-9720664.
    Department of Computer Science, Dartmouth College, Hanover, NH.
    {\tt ney@cs.dartmouth.edu}
  }
}

\pagestyle{myheadings}
\markboth{}{}


\date{}

\maketitle

\thispagestyle{empty}
\begin{abstract}
  \small\baselineskip=10pt
  \parskip=1pt
  
  We study the general (non-metric) facility-location and weighted $k$-medians
  problems, as well as the fractional facility-location and $k$-medians
  problems.  We describe a natural randomized rounding scheme and use it to
  derive approximation algorithms for all of these problems.
  
  For facility location and weighted $k$-medians, the respective algorithms are
  polynomial-time $[H_\Delta k + d]$- and
  $[(1+\epsilon)d;\ln(n+n/\epsilon)k]$-approximation algorithms.  These
  performance guarantees improve on the best previous performance guarantees,
  due respectively to Hochbaum (1982) and Lin and Vitter (1992).  For
  fractional $k$-medians, the algorithm is a new, Lagrangian-relaxation,
  $[(1+\epsilon)d,(1+\epsilon)k]$-approximation algorithm.  It runs in
  $O(k\ln(n/\epsilon)/\epsilon^2)$ linear-time iterations.
  
  For fractional facilities-location (a generalization of fractional weighted
  set cover), the algorithm is a Lagrangian-relaxation,
  $[(1+\epsilon)k]$-approximation algorithm.  It runs in $O(n
  \ln(n)/\epsilon^2)$ linear-time iterations and is essentially the same as an
  unpublished Lagrangian-relaxation algorithm due to Garg (1998).  By recasting
  his analysis probabilistically and abstracting it, we obtain an interesting
  (and as far as we know new) probabilistic bound that may be of independent
  interest.  We call it the {\em Chernoff-Wald} bound.

\end{abstract}
  
\section{Problem definitions}
The input to the {\bf weighted set cover problem} is a collection of sets,
where each set $s$ is given a cost $\cost(s)\in \Rp$.  The goal is to choose a
cover (a collection of sets containing all elements) of minimum total cost.

The (uncapacitated) {\bf facility-location problem} is a generalization of
weighted set cover in which each set $f$ (called a ``facility'') and element
$c$ (called a ``customer'') are given a distance $\dist(f,c)\in \Rp\cup\{\infty\}$.  The goal is to
choose a set of facilities $F$ minimizing $\cost(F)+\dist(F)$, where
$\cost(F)$, the {\em facility cost of $F$}, is $\sum_{f\in F} \cost(f)$ and
$\dist(F)$, the {\em assignment cost of $F$}, is
\(\sum_{c}\min_{f\in{}F}\dist(f,c)\).

\begin{figure}
\vspace{-1em}
    \begin{linearprogram}{minimize$_x$}{d+k}
      \cost(x) & \le & k  \\
      \dist(x) & \le & d \\
      \sum_f x(f,c) & = & 1 & (\forall c) \\
      x(f,c) & \le & x(f) & (\forall f,c) \\
      x(f,c) & \ge & 0 & (\forall f,c) \\
      x(f) & \in & \{0,1\} & (\forall f)
      \label{IP}
    \end{linearprogram}
\vspace{-1em}
  \caption{\sf The facility-location IP.
    Above $d$, $k$, $x(f)$ and $x(f,c)$ are variables, $\dist(x)$, the {\em
      assignment cost of $x$}, is $\sum_{fc} x(f,c)\dist(f,c)$, and $\cost(x)$,
    the {\em facility cost of $x$}, is $\sum_f x(f)\cost(f)$.
    }
  \label{FLIP}
\vspace{-1em}
\end{figure}

Fig.~\ref{FLIP} shows the standard integer programming formulation of the
problem --- the {\em facility-location IP} \cite[p.~8]{NW}.
The facility-location {\em  linear} program (LP)
is the same except without the constraint ``$x(f)  \in \{0,1\}$''.
A {\em fractional} solution is a feasible solution to the LP.
{\em Fractional} facility location is the problem of solving the LP.

The {\bf weighted $k$-medians problem} is the same as the facility
location problem except for the following: a positive real number $k$ is given
as input, and the goal is to choose a subset $F$ of facilities minimizing
$\dist(F)$ subject to the constraint $\cost(F) \le k$.  The standard integer
programming formulation is the {\em $k$-medians IP}, which differs from the
IP in Fig.~\ref{FLIP} only in that $k$ is given, not a variable, and the
objective function is $d$ instead of $d+k$.  The {\em (unweighted)} $k$-medians
problem is the special case when each $\cost(f)=1$.  For the {\em fractional}
$k$-medians problem, the input is the same; the goal is to solve the linear
program obtained by removing the constraint ``$x(f) \in \{0,1\}$'' from the
$k$-medians IP.

We take the ``size'' of each of the above problems to be the number
of pairs $(f,c)$ such that $\dist(f,c) < \infty$.  We assume the size is at
least the number of customers and facilities.

By an $[\alpha(d);\beta(k)]$-{\bf approximation algorithm} for $k$-medians, we
mean an algorithm that, given a problem instance for which there exists a
fractional solution of assignment cost $d$ and facility cost $k$, produces a
solution of assignment cost at most $\alpha(d)$ and facility cost at most
$\beta(k)$.  We use similar non-standard notations for facility location, set cover, and
$k$-set cover.  For instance, by a $[d+2k]$-approximation algorithm for
facility location, we mean an algorithm that, given an instance for which there
exists a fractional solution of assignment cost $d$ and facility cost $k$,
produces a solution for which the assignment cost plus the facility cost is at
most $d+2k$.

\newpage
\section{Background}
In the mid 1970's Johnson and Lovasz gave a greedy $[H_\Delta k]$-approximation
algorithm for unweighted set cover \cite{Johnson74,Lovasz75}.
In 1979 Chvatal generalized it to a
$[H_\Delta k]$-approximation algorithm for weighted set cover \cite{Chvatal79}.

In 1982 Hochbaum gave a greedy $[H_\Delta (d+k)]$-approximation algorithm for
the uncapacitated facility-location problem by an implicit reduction to the
weighted set-cover problem \cite{Hochbaum82}.\footnote{%
  Hochbaum's reduction is easy to adapt in order to reduce unweighted
  $k$-medians to a variant of set cover that we call $k$-set cover.  The
  reduction also extends naturally to the fractional problems.  (See appendix
  for details.)  This shows a loose equivalence between facility location and
  weighted set cover, and between $k$-medians and $k$-set cover.  However,
  Hochbaum's reduction does not preserve the distinction between facility costs
  and assignment costs.  For this reason, we work with the facility location
  and $k$-medians representations.  The algorithms and analyses we give for
  facility location easily imply corresponding results for weighted set cover,
  and the results for $k$-medians are straightforward to adapt to $k$-set
  cover.  }
Above $\Delta$ is at most the maximum, over all facilities $f$, of the number
of customers $c$ such that $\dist(f,c)<\infty$.

In 1992 Lin and Vitter gave a polynomial-time
$[(1+\epsilon)d;(1+1/\epsilon)(\ln{}n+1)k]$-approximation algorithm for the
$k$-medians problem \cite{LinV92}.  (Here $\epsilon>0$ is an input parameter
that tunes the tradeoff between the two criteria and $n$ is the number of
customers.)  Their algorithm combines a greedy algorithm with a technique they
call {\em filtering}.

In 1994 Plotkin, Shmoys, and Tardos (PST) gave Lagrangian-relaxation algorithms
for general packing and covering problems \cite{PlotkinST94}.  As a special
case, their algorithms imply a $[(1+\epsilon)k]$-approximation algorithm for
fractional set cover that runs in $O(k\ln(n)/\epsilon^2)$ linear-time
iterations.  Here $n$ is the number of elements.

In 1998 Garg generalized and simplified the PST set cover algorithm to obtain a
$[(1+\epsilon)k]$-approximation algorithm for fractional {\em weighted} set
cover \cite{GargU98}.
Garg's algorithm runs in $O(n \ln(n)/\epsilon^2)$ linear-time iterations.
By Hochbaum's reduction, one can use Garg's algorithm as a
 $[(1+\epsilon)k]$-approximation algorithm for fractional {\em facility location}.
The running time is the same, where $n$ is the number of customers.

Recent work has focused on {\em metric} $k$-medians and facility location
problems.  In the metric versions, the distance function is assumed to satisfy
the triangle inequality.  For example, $[O(d+k)]$-approximation algorithms have
recently been shown for the metric facility-location problem
\cite{GuhaK1998,ShmyosTA1997}.  Charikar, Guha, Tardos and Shmoys \cite{CGTS99}
recently gave an $[O(d);k]$-approximation algorithm for metric $k$-medians.
Many of these algorithm first solve the fractional problems and then round the
fractional solutions.

\section{Results}
\begin{figure*}[t]
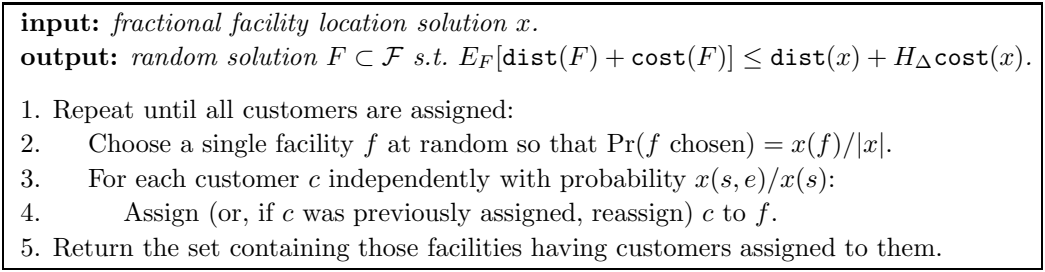

  \figskip
  \begin{center}\hspace*{-0.1in}
\\\framebox{
      \parbox{5.3in}{
        \setlength{\baselineskip}{13pt}
        \inp{fractional facility location solution $x$.}
        \out{random solution $F\subset \cal F$ s.t.\
          $E_F[\dist(F)+\cost(F)] \le \dist( x) + H_{\Delta} \cost( x)$.}
        \setlength{\baselineskip}{13pt}
        1. Repeat until all customers are assigned:
        \\2. \tab Choose a single facility $f$ at random
        so that $\Pr(f \textrm{ chosen}) =  x(f)/|x|$.
        \\3. \tab For each customer $c$ independently with probability
        $ x(s,e)/ x(s)$:
        \\4. \tab \tab Assign (or, if $c$ was previously assigned, reassign) $c$ to $f$.
        \\5. Return the set containing those facilities having customers assigned to them.
        }}
  \end{center}
  \caption{\sf Facility-location rounding scheme.   Note $|x| \doteq \sum_s x(s)$.}
  \label{FLRS}
\end{figure*}

Fig.~\ref{FLRS} shows the simple randomized rounding scheme at the center of
all our results.  With minor variations (e.g.~Fig.~\ref{FKMRS}), this rounding
scheme can be used as the basis for approximation algorithms for set cover,
weighted set cover, facility location, and $k$-medians, and as the basis for
Lagrangian-relaxation algorithms for the fractional variants of these
problems.

Although essentially the same rounding scheme suffices for each of these
problems, the respective probabilistic {\em analyses} require different (albeit
standard) techniques in each case.  For set cover, a simple direct analysis
suffices \cite{Young95}.  For {\em weighted} set cover, facility location, and
$k$-medians, a basic probabilistic lemma called {\em Wald's inequality} is
necessary.  For {\em fractional} set cover and $k$-medians, the
analysis rests on the {\em Chernoff} bound.  For fractional {\em weighted} set
cover and facility location, the analysis is a simple application of 
what we call the {\em Chernoff-Wald} bound.

For each problem, we apply the method of conditional probabilities to the
rounding scheme in order to derive a corresponding approximation algorithm.
The structure of each resulting algorithm, being closely tied to the underlying
probabilistic analysis, ends up differing substantially from problem to problem.

For facility location, the resulting algorithm (Fig.~\ref{FLRS}) is a
randomized rounding, polynomial-time $[d+H_{\Delta}k]$-approximation algorithm.
The performance guarantee improves over Hochbaum's algorithm with respect to
the assignment costs.

For weighted $k$-medians, the resulting algorithm (Fig.~\ref{GKMA}) is a greedy
$[(1+\epsilon)d;\ln(n+n/\epsilon)k]$-approximation algorithm.  In comparison to
Lin and Vitter's algorithm, the performance ratio with respect to the facility
costs is better by a factor of roughly $1/\epsilon$.

For {\em fractional} $k$-medians, the algorithm (Fig.~\ref{FA}) is a
$[(1+\epsilon)d,(1+\epsilon)k]$-approximation algorithm.  It is a
Lagrangian-relaxation algorithm and runs in $O(k\ln(n/\epsilon)/\epsilon^2)$
linear-time iterations.  This is a factor of $n$ faster than the best bound we
can show by applying the general algorithm of Plotkin, Shmoys and Tardos.

Finally, for fractional facility location, the algorithm (see Fig.~\ref{FFLA}) is a
$[(1+\epsilon)(d+k)]$-approximation, Lagrangian-relaxation algorithm.
It runs in at most $O(n\ln(n)/\epsilon^2)$ iterations, where  each iteration requires
time linear in the input size times $\ln n$.  This algorithm is the same
as the unpublished fractional weighted set-cover algorithm due to Garg \cite{GargU98}.

The main interest of this last result is not that we improve Garg's algorithm (we
don't!), but that we recast and abstract Garg's analysis to obtain an apparently new (?)
probabilistic bound --- we call it the Chernoff-Wald bound --- that may be of
general interest for {\em probabilistic} applications.\footnote{%
  The Chernoff-Wald bound also plays a central role in the randomized-rounding
  interpretation of the Garg and Konemann's recent multicommodity-flow
  algorithm \cite{GargK98}.  This and the general connection between randomized
  rounding and greedy/Lagrangian-relaxation algorithms are explored in depth in
  the journal version of \cite{Young95}, which as of October 1999 is still being written.}

A basic contribution of this work is to identify and abstract out (using the
probabilistic method) common techniques underlying the design and analysis of
Lagrangian-relaxation and greedy approximation algorithms.

\section{Wald's inequality and Chernoff-Wald bound}\label{cwsec}

Before we state and prove Wald's inequality and the Chernoff-Wald bound, we
give some simple examples.  Suppose we perform repeated trials of a random
experiment in which we roll a 6-sided die and flip a fair coin.  We stop as
soon as the total of the numbers rolled exceeds $3494$.  Let $T$ be the number
of trials.  Let $D\le 3500$ be the total of the numbers rolled.  Let $H$ be the
number of flips that come up heads.  Since the expectation of the number in
each roll is $3.5$, Wald's implies $E[D] = 3.5 E[T]$, which implies $E[T] \le
1000$.  Since the probability of a head in each coin flip is $0.5$, Wald's
implies $E[H] = 0.5 E[T]$.  Thus we can conclude $E[H]\le500$.

Now modify the experiment so that in each trial, each person {\em in a group of
  50} flips their own fair coin.  Let $M$ be the maximum number of heads any
person gets.  Then Chernoff-Wald states that $E[M] \le (1+\epsilon)500$ for
$\epsilon\approx 0.128$ (so $(1+\epsilon)500 \approx 564$).

If we were to modify the experiment so that the number of trials $T$ was set at
$1000$, the Chernoff-Wald bound would give the same conclusion, but in that
case the {\em Chernoff} bound would also imply (for the same $\epsilon$)
that $\Pr[M \ge(1+\epsilon)500]< 1$

\newcommand{\walds}{
  Let $T\in\Np$ be a random variable with $E[T]< \infty$
  and let $X_1,X_2,\ldots$ be a sequence of random variables.
  Let $\mu,c \in \R$.  If
  \[E[X_{t} \giv T \ge t] \,\le\, \mu ~~(\mbox{for } t \ge 1)\]
  and $X_t \le c$ (for $t=1,\ldots,T$), then
  \[E(X_1+X_2+\cdots+X_T) \, \le \, \mu E(T).\]
  The claim also holds if each ``$\le$'' is replaced by ``$\ge$''.
}
{\samepage
\begin{lemma}[\bf Wald's inequality]\label{walds}
  \walds
\end{lemma}
}

The condition ``$X_t \le c$'' is necessary. Consider choosing each $X_t$
randomly to be $\pm 2^t$ and letting $T = \min\{t : X_t > 0\}$.  Then $E[X_t |
T \ge t] = 0$, so taking $\mu=0$, all conditions for the theorem except ``$X_t
< c$'' are met.  But the conclusion $E[X_1+X_2+\cdots+X_T]\le 0$ does not hold,
because $X_1+X_2+\cdots+X_T = 1$.

The proof is just an adaptation of the proof of Wald's {\em equation}
\cite[p.\ 370]{BE98}.  The reader can skip it on first reading.

\begin{proof}
  W.l.o.g.\ assume $\mu=0$, otherwise apply the change of variables
  $X'_t = X_t - \mu$ before proceeding.
  Define $Y_t \doteq X_1+X_2+\cdots+X_t$.
  If $E(Y_T) = -\infty$ then the claim clearly holds.  Otherwise,
  \begin{eqnarray*}
    E(Y_T) &=&\sum_{t=1}^\infty \Pr(T=t) E(Y_t \giv T=t)
    \\ &=&\sum_{t=1}^\infty \sum_{s=1}^t \Pr(T=t)E(X_s \giv T=t).
  \end{eqnarray*}
  The sum of the positive terms above is at most
  $\sum_t \sum_{s\le t} \Pr(T=t)c = \sum_t \Pr(T=t) ct = E[cT] < \infty$.
  Thus, the double sum is absolutely convergent so
  \begin{eqnarray*}
    \phantom{E(Y_T)}
    & = &\sum_{s=1}^\infty \sum_{t=s}^\infty \Pr(T=t)E(X_s \giv T=t)
    \\&= &\sum_{s=1}^\infty \Pr(T\ge s)E(X_s \giv T \ge s)
  \end{eqnarray*}
  This establishes the claim because $E(X_s\giv T\ge s) \le 0$.
  The claim with ``$\le$'''s replaced by ``$\ge$'''s follows
  via the change of variables $X'_t \doteq -X_t$,
  $\mu' = -\mu$, and $c'\doteq-c$. ~\hfill\qed
\end{proof}

In many applications of Wald's, the random variables $X_1,X_2,\ldots$ will be
independent, and $T$ will be a {\em stopping time} for $\{X_t\}$ ---
a random variable in $\Np$ such that the event ``$T=t$'' is independent of
$\{X_{t+1},X_{t+2},\ldots\}$.   In this case the following companion lemma
facilitates the application of Wald's inequality:
\begin{lemma}\label{stoptime1}
  Let $X_1,X_2,\ldots$ be a sequence of independent random variables
  and let $T$ be a stopping time for the sequence.
  Then $E[X_t \giv T\ge t] = E[X_t]$.
\end{lemma}
\begin{proof}
  Because $X_t$ is independent of $X_1,X_2,\ldots,X_{t-1}$, and the event
  ``$T\ge t$'' (i.e.\ $T\not\in\{1,2,\ldots,t-1\}$) is determined by the values
  of $X_1,X_2,\ldots,X_{t-1}$, it follows that $X_t$ is independent of the
  event ``$T\ge t$''.  \hfill\qed
\end{proof}

Here is a statement of a standard Chernoff bound.
For a proof see e.g.~\cite{Raghavan88,Young95}.
\newpage

Let
$\chern(\epsilon) \doteq(1+\epsilon)\ln(1+\epsilon)-\epsilon\ge\epsilon\ln(1+\epsilon)/2$.
\\For $\epsilon \le 1$, $\chern(\epsilon) \ge \epsilon^2/3$ and $\chern(-\epsilon)\ge \epsilon^2/2$.
\begin{lemma}[{\bf Chernoff Bound} \cite{Raghavan88}]\label{ChernoffLower}~
  \\Let $X_1,X_2,\ldots,X_k$ be a sequence of independent random variables in $[0,1]$
  with $E(\sum_i  X_i) \ge \mu > 0$.
  Let $\epsilon>0$.
  \\Then
  \(\Pr\left[ \sum_i X_i \ge \mu(1+\epsilon)\right] < \exp(-\chern(\epsilon) \mu)\).
  \\For $\epsilon < 1$,
  \(\Pr\left[ \sum_i X_i \le \mu(1-\epsilon)\right] < \exp(-\chern(-\epsilon) \mu).\)
\end{lemma}

\newcommand{\chernoffWald}{
  For each $i=1,2,\ldots,m$, let $X_{i1},X_{i2},\ldots$ be a sequence of 
  random variables such that $0\le X_{it}\le 1$.
  \\Let $T\in\Np$ be a random variable with $E[T] < \infty$.
  \\Let $M = \max_{i=1}^m X_{i1}+X_{i2}+\cdots+X_{iT}$.
  Suppose
  \[E[X_{it} \giv T\ge t; \{X_{jl}:j < t, 1\le i\le m\}] \,\le\,  \mu\]
  for all $i,t$ for some $\mu \in \R$.  Let $\epsilon\ge 0$ satisfy
  \[e^{\textstyle-\chern(\epsilon) \max\{\mu E[T],E[M]/(1+\epsilon)\}} \le 1/m.\]
  Then
  \[E[M] \, \le \, (1+\epsilon) \mu E[T].\]

  The claim also holds with the following replacements:
  ``$\min_i$'' for   ``$\max_i$'';
  ``$\ge \mu$'' for   ``$\le \mu$'';
  ``$E[M]  \ge $'' for   ``$E[M] \le $'';
  and, for each ``$\epsilon$'', ``$-\epsilon$''
  (except in ``$\epsilon\ge 0$'').
}

{\samepage
\begin{theorem}[\bf Chernoff-Wald Bound]\label{chernoff-wald}~
\\  \chernoffWald
\end{theorem}
}

In the case when $T$ is constant, the Chernoff bound 
implies $\Pr[M \ge (1+\epsilon)\mu T] < 1$ for $\epsilon$ as above.

The first-time reader can skip the following proof.
\begin{proof}
  For $t=0,\ldots,T$, let $Y_{it} \doteq X_{i1}+X_{i2}+\cdots+X_{it}$ ($i=1,\ldots,m$)
  and
  \[Z_t \doteq \log_{1+\epsilon} \sum_{i=1}^m (1+\epsilon)^{Y_{it}}.\]
  Note that for each $i$,  $Z_T \ge \log_{1+\epsilon} (1+\epsilon)^{Y_{iT}} = Y_{iT}$.
  Thus $M \le Z_T$.  
  We use Wald's inequality to bound $E[Z_T]$.
  Fix any $t>0$.  Let $y_{it} = (1+\epsilon)^{Y_{i,t-1}}$.
  Then
  \begin{eqnarray*}
    Z_t-Z_{t-1} & = &
    \log_{1+\epsilon} \frac{ \sum_i y_{it}(1+\epsilon)^{X_{it}}}{\sum_i y_{it}}
    \\& \le &
    \log_{1+\epsilon} \frac{ \sum_i y_{it}(1+\epsilon X_{it})}{\sum_i y_{it}}
    \\& = & 
    \log_{1+\epsilon} \left[1+\epsilon \frac{\sum_iy_{it}X_{it}}{\sum_i y_{it}}\right]
    \\& < & 
    \frac{\epsilon}{\ln(1+\epsilon)} \frac{ \sum_i y_{it}X_{it}}{\sum_i y_{it}}.
  \end{eqnarray*}
  The first inequality follows from $(1+\epsilon)^z \le 1+\epsilon z$ for  $0\le z\le 1$.
  The last inequality follows from
  $\log_{1+\epsilon} 1+z =  \ln (1+z)/\ln(1+\epsilon) < z/\ln(1+\epsilon)$ for $z\neq 0$.
  Thus, conditioned on the event $T\ge t$ and the values of $Y_{i,t-1} (1\le i \le m)$,
  \[E[Z_t-Z_{t-1}] \,<\,
  \frac{\epsilon}{\ln(1+\epsilon)} \frac{ \sum_i y_{it}\, \mu }{\sum_i y_{it}}.
  \]
  This implies that
  $E[Z_t - Z_{t-1} | T\ge t] < {\mu \epsilon}/{\ln(1+\epsilon)}$.
  Using $Z_0 = \log_{1+\epsilon} m$
  and Wald's inequality,
  \[E[M] \le E[Z_T] < \log_{1+\epsilon} m \,+\, E[T]\mu \epsilon/\ln(1+\epsilon).\]
  By algebra, the above
  together with the assumption on $\epsilon$
  imply that $E[M] \le (1+\epsilon) \mu E[T]$.

  The proof of the claim for the minimum 
  is essentially the same, with each ``$\epsilon$'' replaced by ``$-\epsilon$''
  and reversals of appropriate inequalities.
  In verifying this, note that $\log_{1-\epsilon}$ is a decreasing function. \hfill\qed
\end{proof}

In many applications of Chernoff-Wald, each 
$X_{it}$ will be independent of $\{X_{j\ell }: \ell < t, 1\le j \le m\}$,
and $T$ will be a stopping time for $\{X_{it}\}$ ---
a random variable in $\Np$ such for any $t$ the event ``$T = t$''
is independent of $\{X_{i\ell} : \ell > t, 1\le i \le m\}$.
Then by an argument similar to the proof of Lemma~\ref{stoptime1}, we have:
\begin{lemma}~\label{stoptime2}
  For $1\le i\le m$, let $X_{i1},X_{i2},\ldots$ be a sequence of random
  variables such that each $X_{it}$ is independent of
  $\{X_{j\ell} : \ell < t, 1\le j \le m\}$.
  Let $T$ be a stopping time for $\{X_{it}\}$.
  Then
  \\$E[X_{it} \giv T\ge t; \{X_{jl}:j < t, 1\le i\le m\}] = E[X_{it}]$.
\end{lemma}

\medskip
\section{Randomized rounding for facility location. }
We use Wald's inequality to analyze a natural randomized rounding scheme
for facility location.
\begin{guarantee}\label{FLRSG}
  Let $F$ be the output of the facility-location rounding scheme in
  Fig.~\ref{FLRS} given input $x$.  Let $\Delta_{fx}$ be the number of
  customers $c$ such that $x(f,c) >0$.
  Then $E_F[\dist(F)+\cost(F)]$ is at most
  $\dist(x) + \sum_f \cost(f) x(f) H_{\Delta_{fx}}$.
\end{guarantee}
\begin{proof}
  Observe the following basic facts about each iteration of the outer loop:
  \begin{enumerate}
  \item The probability that a given customer $c$ is assigned to a particular
    facility $f$ in this iteration is \(x(f,c)/|x|.\)
  \item The probability that $c$ is assigned to {\em some} facility 
    is \(\sum_f  x(f,c)/|x| \ge 1/|x|.\) 
  \item  {\em Given} that $c$ is assigned, the probability of it being assigned 
    to a particular $f$ is $ x(f,c)$.
  \end{enumerate}
  To bound $E[\dist(F)]$, it suffices to bound the expected cost of the
  assignment chosen by the algorithm.
  For a given pair $(f,c)$, what is the probability that $f$ is assigned to
  $c$?  By the third fact above, this is $x(f,c)$.
  Thus, 
  $E[\dist(F)] \le \sum_{f,c} \Pr(c \mbox{ assigned to } f) \dist(f,c) = \dist(x)$.

  To finish, we show  $\Pr(f \in F) \le H_{\Delta_{fx}}x(f)$.   
  This suffices because it implies that $E[\cost(F)]$ equals
  $\sum_f \Pr(f \in F)\cost(f) \le \sum_f H_{\Delta_{fx}} x(f) \cost(f)$.
  
  Fix a facility $f$.  Call the customers $c$ such that $x(f,c)>0$ the
  ``fractional customers of $f$''.  Let random variable $T$ be the number of
  iterations before all these customers are assigned.
  \\ {\bf claim 1:} {\em $\Pr(f\in F) \le E[T] x(f)/|x|$}.
  \\ {\bf proof:} Define $X_t$ to be the indicator variable for the
  event ``$f$ is first chosen in iteration $t$''.

  As $E[X_t | T \ge t] \le x(f)/|x|$, by Wald's inequality
  $\Pr(f\in F) = E[X_1+X_2+\cdots+X_T] \le E[T]x(f)/|x|$.
  This proves the claim.
  
  Recall that $\Delta_{fx}$ is the number of fractional customers of $f$.
  To finish the proof, it suffices to show:
  \\ {\bf claim 2:} {\em $E[T] \le |x| H_{\Delta_{fx}}$}
  \\ {\bf proof:} Define $u_{t}$ to be the number of fractional
  customers of $f$ not yet assigned  after iteration $t$ ($0 \le t \le T$).
  (Recall $H_0 \doteq 0$ and $H_i \doteq 1+1/2+\cdots+1/i$.) Then provided
  $t \le T$,    $H_{u_{t}} - H_{u_{t+1}}$ is
  \[
    \frac{1}{u_{t}} + \frac{1}{u_{t}-1} + \cdots + \frac{1}{u_{t+1}-1}
    \,\ge\, \frac{u_{t}-u_{t+1}}{u_{t}}.
    \]
    The expectation of the right-hand side is at least $1/|x|$ because each
    customer is assigned in each iteration with probability at least $1/|x|$.
    Since $H_{u_{t}}$ is $H_{\Delta_{fx}}$ when $t=0$ and decreases by at least
    $1/|x|$ in expectation each iteration, by Wald's inequality, it follows
    that $E[H_{\Delta_{fx}} - H_{u_{T}}] \ge E[T]/|x|$.  Since $H_{u_{T}} = 0$,
    the claim follows. \hfill\qed
\end{proof}
The randomized scheme can easily be derandomized.
\begin{corollary}
  There is a polynomial-time $[d + H_{\Delta} k]$-approximation algorithm for
  uncap.\ facility location.
\end{corollary}

\begin{figure*}[t]
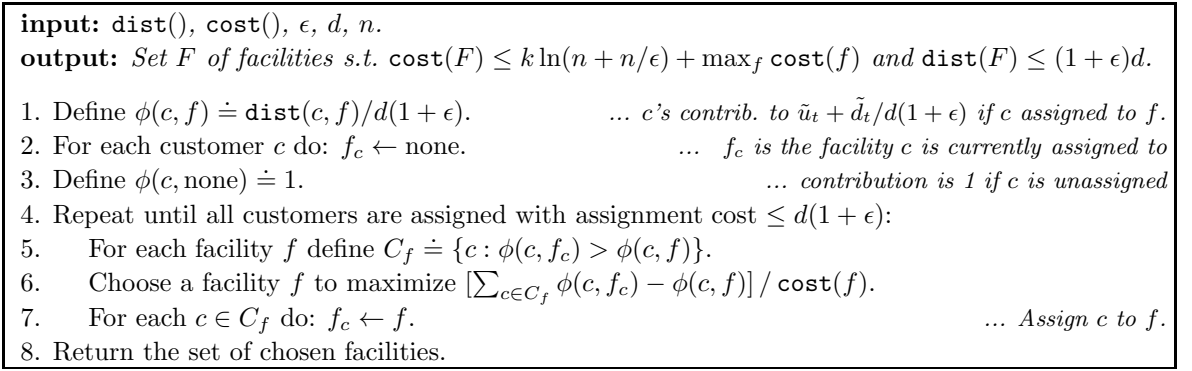

  \figskip
  \begin{center}\hspace*{-0.1in}
\\\framebox{
      \parbox{6in}{
        \setlength{\baselineskip}{13pt}
  \inp{$\dist()$, $\cost()$, $\epsilon$, $d$, $n$.}
  \out{Set $F$ of facilities s.t.\
    $\cost(F) \le k \ln(n+n/\epsilon)+\max_f \cost(f)$
    and $\dist(F) \le (1+ \epsilon)d$.}
    \setlength{\baselineskip}{13pt}
  1. Define $\phi(c,f) \doteq \dist(c,f)/d(1+\epsilon)$.
  \comment{$c$'s contrib.\ to $\rnd u_{t} + \rnd d_{t}/d(1+\epsilon)$
    if $c$ assigned to $f$.}
  \\2. For each customer $c$ do: $f_c \leftarrow \mathrm{none}$.
  \comment{ $f_c$ is the facility $c$ is currently assigned to}
  \\3. Define $\phi(c,\mathrm{none}) \doteq 1$.
  \comment{contribution is 1 if $c$ is unassigned}
  \\4. Repeat until all customers are assigned
  with assignment cost $\le d(1+\epsilon)$:
  \\5. \tab For each facility $f$ define $C_f \doteq \{ c : \phi(c,f_c) > \phi(c,f)\}$.
  \\6. \tab Choose a facility $f$ to maximize
  \([\sum_{c\in C_f} \phi(c,f_c)-\phi(c,f)]\,/\,\cost(f).\)
  \\7. \tab For each $c\in C_f$ do: $f_c \leftarrow f$.
  \comment{Assign $c$ to $f$.}
  \\8. Return the set of chosen facilities.
  }}
  \caption{\sf The greedy $k$-medians algorithm.}
  \label{GKMA}
  \label{WIA}
  \end{center}
  \vspace*{-2ex}
\end{figure*}

\section{Greedy algorithm for weighted $k$-medians. }
The {\em $k$-medians rounding scheme} takes a fractional
$k$-medians solution $(x,k,d)$ and an $\epsilon>0$ and outputs a random
solution $F$.  The scheme is the same as the facility-location
rounding scheme in Fig.~\ref{FLRS} except for the termination condition.
The modified algorithm terminates after the first iteration in which
the facility cost exceeds $k[\ln(n+n/\epsilon)]$ {\bf or} all customers are
assigned with assignment cost less than $d(1+\epsilon)$.
We analyze this rounding scheme using Wald's inequality
and then derandomize it to obtain a greedy algorithm (Fig.~\ref{WIA}).

\begin{guarantee}\label{IRSG}
  Let $F$ be the output of the weighted $k$-medians rounding scheme given input
  $x$.  
  Then $\cost(F) \le k\ln(n+n/\epsilon)+\max_f\cost(f)$ and with
  positive probability $\dist(F) < d(1+ \epsilon)$.
\end{guarantee}
\begin{proof}
  The bound on $\cost(F)$ always holds due to the termination condition of the
  algorithm.
  
  Let random variable $T$ be the number of iterations of the rounding
  scheme.  Let random variable $u_t$ be the number of not-yet-assigned
  customers at the end of round $t$ ($0\le t\le T$).  By fact 2 in the
  proof of Guarantee~\ref{FLRSG},
  $E[u_t | u_{t-1} \wedge T \ge t] = (1-1/|x|) u_{t-1}$.
  
  Define random variable $d_t$ to be the total cost of the current (partial)
  assignment of customers to facilities at the end of round $t$.  Because each
  customer is reassigned with probability $1/|x|$ in each round, it is not to
  hard to show that \(E[d_t \giv d_{t-1} \wedge T \ge t]=(1-1/|x|)d_{t-1}+d/|x|.\)
  
  Define random variable $c_t$ to be the total cost of the facilities chosen so
  far at the end of round $t$.  In each iteration, $E[\cost(f)] = k/|x|$, so
  that \(E[c_t \giv c_{t-1} \wedge T \ge t] = c_{t-1} + k/|x|.\)

  Define $\phi_t \doteq c_t/k\,+\,\ln[u_t +  (d_t/d-1)\,/\,(1+\epsilon)].$
  Then using $\ln z \le z-1$ and the two facts established in the preceding three 
  paragraphs, a calculation shows
  \(E[\phi_t-\phi_{t-1} | d_{t-1},u_{t-1},c_{t-1}] \le 0.\)
  By Wald's inequality, this implies
  \(E[\phi_T] \le \phi_0 < \ln n.\)
  Thus, with positive probability,
  \(\phi_T \le \ln n.\)
  Assuming this event occurs, we will show that at the end
  all elements are covered and the assignment cost is not too high.

  If the rounding scheme terminates because
  all elements are covered and the assignment cost is less than $(1+\epsilon)d$,
  then clearly the performance guarantee holds.
  Otherwise the algorithm terminates because
  the facility cost $c_T$ exceeds $\ln(n+n/\epsilon)k$.
  This lower bound on the size and
  the occurrence of the event
  ``$\phi_T \le \ln n$''  imply that  $u_T < 1$  and $d_T < (1+\epsilon)d$. \hfill\qed
\end{proof}

\begin{figure*}
  \figskip
  \begin{center}\hspace*{-0.1in}
\\\framebox{
  \parbox{5.7in}{
    \setlength{\baselineskip}{13pt}
    \inp{fractional $k$-medians solution $(\opt x,k,d)$, $0<\epsilon <1$.}
    \out{random fractional solution $\rnd x$ s.t.\ $\cost(\rnd x) = (1-\epsilon)^{-1} k$
      and  $E[\dist(\rnd x)] \le (1-\epsilon)^{-2}d$.} \setlength{\baselineskip}{13pt}
    1. Choose $N \ge \ln(n/\epsilon)/\chern(-\epsilon)$ s.t.\ $N\,k$ is 
    an integer.  \comment{Recall $\chern(\epsilon) \ge \epsilon\ln(1+\epsilon)$.}
    \\2. For each $f$,$c$ do:  $x(f) \leftarrow x(f,c) \leftarrow x(c) \leftarrow 0$.
  \\3. Repeat $N\,k$ times:
  \\4. \tab Choose a single facility $f$ at random
  so that $\Pr(f \textrm{ chosen}) = \opt x(f)/k$.
  \\5. \tab Increment $x(f)$.
  \\6. \tab For each customer $c$, with probability
  $\opt x(f,c)/\opt x(f)$ do:
  \\7. \tab \tab Increment $x(f,c)$ and $x(c)$.
  \\8. Return $\rnd x$, where $\rnd x \doteq x\,/\,(1-\epsilon)N$.  
  }}
  \end{center}
  \caption{\sf Fractional $k$-medians rounding scheme.
    }
  \label{FKMRS}
\end{figure*}

Next we apply the method of conditional probabilities.
Let $T$, $d_t$, $u_t$, $c_t$, and $\phi_t$ ($0\le t\le T$) be defined as in the 
proof of Guarantee~\ref{IRSG} for the $k$-medians rounding scheme.
That proof showed that $E[\phi_T] < \ln n$,
and that if $\phi_T < \ln n$ then $F$ meets the performance guarantee.

To obtain the greedy algorithm, in each iteration we replace the random choices
by deterministic choices.  Let $\rnd d_t$, $\rnd u_t$, and $\rnd c_t$ denote,
respectively, the assignment cost, number of unassigned elements, and facility
cost at the end of the $t$th iteration of the greedy algorithm
(analogous to $d_t$, $u_t$, and $c_t$ for the randomized algorithm).
The greedy algorithm will make its choices in a way that maintains
the invariant
\[E[\phi_T \giv d_t=\rnd d_t \wedge u_t =\rnd u_t \wedge c_t = \rnd c_t] < \ln n.\]
Note that the expectation above is with respect to the random
experiment.  That is, the invariant says that {\em if}, starting from the
current configuration, the {\em remaining} choices were to be made randomly,
then (in expectation) $\phi$ would end up less than $\ln n$.

Define $\rnd\phi_t \doteq \rnd c_t/k\,+\,\ln[\rnd u_t +  (\rnd d_t/d-1)\,/\,(1+\epsilon)]$
(analogous to $\phi_t$ for the randomized algorithm).
The proof of Guarantee~\ref{IRSG} easily generalizes to show
\(E[\phi_T \giv d_t=\rnd d_t \wedge u_t =\rnd u_t \wedge c_t = \rnd c_t] \le \rnd\phi_t.\)
Thus, it suffices to maintain the invariant $\rnd\phi_t < \ln n$.
Since $\rnd\phi_0 = \phi_0 < \ln n$, the invariant holds initially.

During each iteration $t$, the algorithm chooses a facility $f$ and assigns it a
set of customers $C$ so that $\rnd \phi_t \le \rnd \phi_{t-1}$.
A calculation shows
$\rnd \phi_t - \rnd \phi_{t-1}$ is less than
\\
\[
\frac{\cost(f)}{k} -
\frac{\rnd u_{t-1} - \rnd u_{t} + (\rnd d_{t-1}/d - \rnd d_{t}/d)/(1+\epsilon)}%
{\rnd u_{t-1} + (\rnd d_{t-1}/d-1)/(1+\epsilon)}.
\]
It suffices to choose $f$ and $C$ so that the above is non-positive.
Whatever $\rnd u_{t-1}$, $\rnd c_{t-1}$, and $\rnd d_{t-1}$ 
are, if $f$ and $C$ are chosen randomly,
then the expectation of the above is zero. 
Thus, there is {\em some} choice of $f$ and $C$
which makes it non-positive.
Thus it suffices to choose $f$ and $F$ to maximize
\[\frac{\rnd u_{t-1} - \rnd u_{t}
  + (\rnd d_{t-1}/d - \rnd d_{t}/d)/(1+\epsilon)}{\cost(f)}.\]
The algorithm considers each facility $f$;
for each $f$, it determines the best set $C$ of customers to assign.
The algorithm is shown in Fig.~\ref{WIA}.

The termination condition in the algorithm differs from the one in the rounding
scheme, but the modified termination condition suffices because it follows from
the analysis that whatever $k$ is, the algorithm will terminate no later than
the first iteration such that the facility cost exceeds $k \ln(n+n\epsilon)$.
By the derivation,
\begin{guarantee}\label{WIAG}
  Given $\epsilon$,
  and $d$ such that a fractional solution of cost $k$ and assignment cost at
  most $d$ exists,
  the greedy weighted $k$-medians algorithm (Fig.~\ref{WIA}) returns a
  solution $F$ 
  such that
  $\dist(F) \le (1+ \epsilon)d$
  and
  $\cost(F) \le k \ln(n+n/\epsilon)+\max_f \cost(f)$.
\end{guarantee}
Without loss of generality (since we are approximately solving the IP)
$\max_f \cost(f) \le k$.  For the unweighted
problem, the number of iterations is $O(k\ln(n/\epsilon))$.  No facility is
chosen twice, so the number of iterations is always at most $m$, the number of
facilities.
\begin{corollary}  Let $0 < \epsilon < 1$.
  The weighted $k$-medians problem has
  a $[(1+\epsilon)d;(1+\ln(n+n/\epsilon)k)]$-approximation algorithm
  that runs in $O(m)$ linear-time iterations, or
  $O(k\ln(n/\epsilon))$ iterations for the unweighted problem.
\end{corollary}

\section{Lagrangian relaxation for $k$-medians. }
In this section we derive and analyze a Lagrangian-relaxation,
$[(1+\epsilon)^2d;(1+\epsilon)k]$-approximation
algorithm for the fractional unweighted $k$-medians problem.
The rounding scheme is
shown in Fig.~\ref{FKMRS}.
We use the Chernoff and Markov bounds to bound the
probability of failure.  
\begin{guarantee}\label{FRSG}
  Let $\rnd x$ be the output of the fractional $k$-medians rounding scheme.
  Then with positive probability $\rnd x$ has
  $\cost(\rnd x) \le (1-\epsilon)^{-1}k$ and
  $\dist(\rnd x) \le (1-\epsilon)^{-2}d$.
\end{guarantee}
\begin{proof}
  Recall $k=\cost(\opt x) = |\opt x|$ and $d=\dist(\opt x)$.
  The bound on the cost always holds, because each of the $kN$ iterations
  adds 1 to $\cost(x)$.
  It remains to show that with positive probability, after the final iteration,
  $\dist(x) \le (1-\epsilon)^{-1}Nd$
  and each $x(c) \ge (1-\epsilon)N$.

  Since each iteration increases $\dist(x)$ by $d/k$ in expectation,
  finally $E[\dist(x)] \le (Nk)d/k = dN$.
  By the Markov bound, $\Pr[\dist(x) \ge dN/(1-\epsilon)] \le 1-\epsilon$.

  For any customer $c$, $x(c)$ is the sum of $kN$ independent 0-1 random
  variables each with expectation at least $1/k$, so by the Chernoff bound,
  $\Pr[x(c) \le (1-\epsilon)N] < \exp(-\chern(-\epsilon) N)$, which is
  at most $\epsilon/n$ by the choice of $N$.

  By the naive union bound,
  $\Pr[\dist(x) \ge dN/(1-\epsilon) \,\vee\, \min_c x(c) \le (1-\epsilon)N]
  < (1-\epsilon) + n(\epsilon/n) = 1$. \hfill\qed
\end{proof}

\begin{figure*}
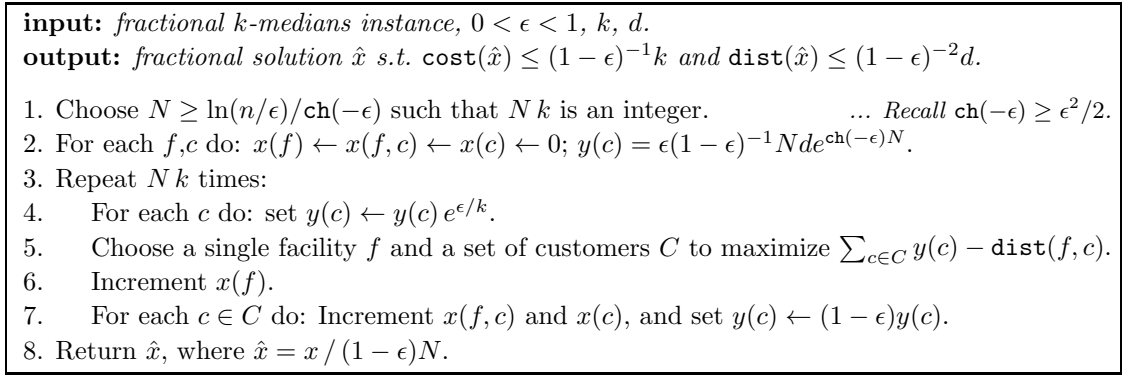

  \figskip
  \begin{center}\hspace*{-0.1in}
  \\\framebox{
          \parbox{5.7in}{
        \setlength{\baselineskip}{13pt}
  \inp{fractional $k$-medians instance, $0 < \epsilon < 1$, $k$, $d$.}
  \out{fractional solution $\dtm x$ s.t.\
    $\cost(\dtm x) \le (1-\epsilon)^{-1}k$ and
    $\dist(\dtm x) \le (1-\epsilon)^{-2}d$.
    }
        \setlength{\baselineskip}{13pt}
  1. Choose $N \ge \ln(n/\epsilon)/\chern(-\epsilon)$ such that $N\,k$ is an integer.
  \comment{Recall $\chern(-\epsilon) \ge \epsilon^2/2$.}
  \\2. For each $f$,$c$ do:  $x(f) \leftarrow x(f,c) \leftarrow x(c) \leftarrow 0$; 
  $y(c) = \epsilon (1-\epsilon)^{-1} N d e^{\chern(-\epsilon)N}$.
  \\3. Repeat $N\,k$ times:
  \\4. \tab For each $c$ do: set $y(c) \leftarrow y(c)\,e^{\epsilon/k}$.
  \\5. \tab Choose a single facility $f$ and a set of customers $C$ to maximize
  \(\sum_{c\in C} y(c) - \dist(f,c).\)
  \\6. \tab Increment $x(f)$.
  \\7. \tab For each $c\in C$ do: Increment $x(f,c)$ and $x(c)$, and set $y(c)\leftarrow(1-\epsilon)y(c)$.
  \\8. Return $\dtm x$, where $\dtm x = x\,/\,(1-\epsilon)N$.  
}}\end{center}
  \caption{\sf Lagrangian-relaxation algorithm for fractional $k$-medians.
    In step 4, it suffices to choose the best set of the form
    $C=\{c : y(c) > \dist(f,c)\}$ for some $f$.
    This can be done in linear time.
    }
  \label{FA}
\end{figure*}

Next we sketch how the method of conditional expectations
yields the algorithm shown in Fig.~\ref{FA}.
The proof of Guarantee~\ref{FRSG} implicitly bounds the probability of failure by
the expectation of
\[
\frac{\dist(x)}{dN/(1-\epsilon)}
+
\sum_c \frac{(1-\epsilon)^{x(c)}}{(1-\epsilon)^{(1-\epsilon)N}}\]
and it shows that the expectation is less than 1.
An upper bound (called a ``pessimistic estimator'' \cite{Raghavan88})
of the conditional expectation of the final value of the above,
given the current value of $x$ and the number $t$ of remaining iterations, is
\[
\pe(x,t) \doteq
\frac{\dist(x) + t\,d/k}{dN/(1-\epsilon)}
+
\sum_c \frac{(1-\epsilon)^{x(c)}e^{-t\,\epsilon/k}
}{(1-\epsilon)^{(1-\epsilon)N}}\]
The algorithm chooses $f$ and $C$ in each iteration
in order to minimize the increase in the above quantity,
which consequently stays less than 1.

\begin{guarantee}\label{FAG}
  The fractional $k$-medians algorithm
  in Fig.~\ref{FA}
  returns a fractional solution $\dtm x$ having
  $\cost(\dtm x) \le (1-\epsilon)^{-1}k$ and
  $\dist(\dtm x) \le (1-\epsilon)^{-2}d$.
\end{guarantee}
\begin{proof} (Sketch.)
  Let $\pe$ be as defined above.
  The algorithm maintains the invariant that with $t$ iterations remaining,
  $\pe(x,t) < 1$.
  It is straightforward to verify that the invariant is initially true,
  and that if is true at the end, then the performance guarantee is met.
  We verify that the invariant is maintained at each step.
  The increase in $\pe$ in a single iteration is proportional to
  \[\sum_{c\in C} \dist(f,c) - \frac{d}{k} + \sum_c y(c)
  e^{\epsilon/k}(1-\epsilon [c\in C]) \,-\, \sum_c y(c)\]
  where $y(c) = \alpha\, (1-\epsilon)^{x(c)}$ before the iteration
  for suitably chosen scalar $\alpha\ge 0$.
  If $f$ and $C$ were chosen randomly as in the rounding scheme,
  the expectation of the above would be at most 0.
  The choice made by the algorithm minimizes the above quantity,
  therefore the algorithm maintains the invariant. \hfill\qed
\end{proof}

\begin{corollary}\label{fkm}
  Let $0 < \epsilon < 1$.
  The fractional $k$-medians problem has a $[(1+\epsilon)d,(1+\epsilon)k]$-approximation algorithm
  that runs in $O(k\ln(n/\epsilon)/\epsilon^2)$
  linear-time iterations.
\end{corollary}

\section{Lagrangian relaxation for facility location. }
The rounding scheme for fractional facility location
is the same as the rounding scheme
for fractional $k$-medians in Fig.~\ref{FKMRS},
except for the termination condition.
The modified algorithm terminates
after the first iteration where each $x(c) \ge (1-\epsilon)N$,
where $N$ is chosen to be at least $\ln(n)/\chern(-\epsilon)$
s.t.\ $(1-\epsilon)N$ is an integer.
We use the Chernoff-Wald bound to analyze the scheme.
\begin{guarantee}\label{FFLG}
  Let $\rnd x$ be the output of the fractional facility-location rounding
  scheme as described above.
  Then $\rnd x$ is a fractional solution to the facility location LP
  and $E[\cost(\rnd x) + \dist(\rnd x)] \le (1-\epsilon)^{-1}(d+k)$.
\end{guarantee}
\begin{proof}
  The termination condition ensures that all customers are adequately covered.
  It remains to bound $E[\cost(\rnd x) + \dist(\rnd x)]$.

  Let r.v.~$T$ denote the number of iterations of the rounding scheme.
  In each iteration, the expected increase in
  $\cost(x) + \dist(x)$ is $k/|\opt x| + d/|\opt x|$.
  By Wald's inequality, at termination,  $E[\cost(x) +\dist(x)] \le E[T](k+d)/|\opt x|$.
  It remains to show $E[T] \le N|\opt x|$.
  (Recall that $\rnd x = x/(1-\epsilon)N$.)

  For any customer $c$, the probability that $x(c)$ is incremented in a given
  iteration is at least $1/|\opt x|$, independently of the previous iterations.
  Let r.v.\ $M \doteq \min_c x(c)$ at the end.
  Note that, by the choice of $N$, in fact $M = (1-\epsilon)N$.

  By the Chernoff-Wald bound, $(1-\epsilon)N = E[M] \ge (1-\epsilon)E[T]/|\opt x|$
  (provided $\chern(-\epsilon) \ge \ln(m)(1-\epsilon)/E[M]$,
  which indeed holds by the choice of $N$).
  Rewriting gives $E[T] \le N|\opt x|$. \hfill\qed
\end{proof}

\begin{figure*}
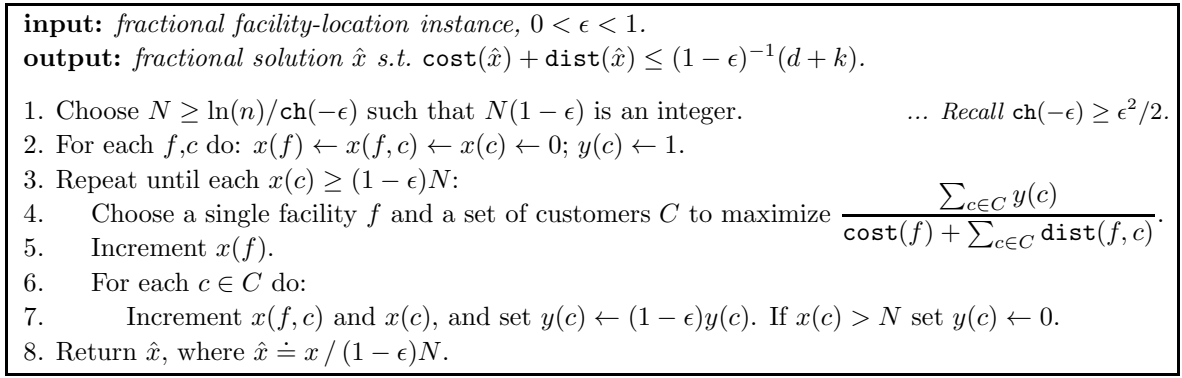

  \figskip
  \begin{center}\hspace*{-0.1in}
  \\\framebox{
          \parbox{6in}{
        \setlength{\baselineskip}{13pt}
  \inp{fractional facility-location instance, $0<\epsilon < 1$.}
  \out{fractional solution $\dtm x$ s.t.\
    $\cost(\dtm x) + \dist(\dtm x) \le (1-\epsilon)^{-1}(d+k)$.
    }
        \setlength{\baselineskip}{13pt}
  1. Choose $N \ge \ln(n)/\chern(-\epsilon)$ such that $N(1-\epsilon)$ is an integer.
  \comment{Recall $\chern(-\epsilon) \ge \epsilon^2/2$.}
  \\2. For each $f$,$c$ do:  $x(f) \leftarrow x(f,c) \leftarrow x(c) \leftarrow 0$; $y(c) \leftarrow 1$.
  \\3. Repeat until each $x(c) \ge (1-\epsilon)N$:
  \\4. \tab Choose a single facility $f$ and a set of customers $C$ to maximize  
  \raisebox{0in}[0in][0in]{\(\displaystyle\frac{\sum_{c\in C} y(c)}{\cost(f)+\sum_{c\in C} \dist(f,c)}.\)}
  \\5. \tab Increment $x(f)$.
  \\6. \tab For each $c\in C$ do:
  \\7. \tab \tab Increment $x(f,c)$ and $x(c)$, and set
  $y(c)\leftarrow(1-\epsilon)y(c)$.
  If $x(c)>N$ set $y(c) \leftarrow 0$.
  \\8. Return $\dtm x$, where $\dtm x \doteq x\,/\,(1-\epsilon)N$.  
  }}\end{center}
  \caption{\sf Lagrangian-relaxation algorithm for fractional facility location.
    In step 4, it suffices to choose the best set of the form
    $C=\{c : y(c)/\dist(f,c) \ge \lambda\}$ for some $\lambda$.
    }
  \label{FFLA}
\end{figure*}

Next we sketch how applying the method of conditional expectations gives the
Lagrangian-relaxation algorithm shown in Fig.~\ref{FFLA}.
Below we let $x_f$ denote the value of $x$ after the final iteration of the
algorithm and $x$ denote the value at the ``current'' iteration.
The analysis of the
rounding scheme shows that $E[\cost(x_f)+\dist(x_f)] \le N\,(k+d)$.  The
conditional expectation of $\cost(x_f)+\dist(x_f)$ at the end, given the current
$x$, is $\cost(x)+\dist(x)+E[t | x](k+d)/|x^*|$, where random variable
$t$ is the number of iterations left.

The proof of Chernoff-Wald, in this context, argues that the quantity
$\rnd M(x) \doteq \log_{1-\epsilon} \sum_c (1-\epsilon)^{x(c)}$ is
$\log_{1-\epsilon}n$
initially, at most $(1-\epsilon)N$ finally, and decreases in expectation at
least $-\epsilon/(|\opt x|\ln(1-\epsilon))$ in each iteration.  An easy
generalization of the argument shows $E[t|x]$ is at most
\(\frac{(1-\epsilon)N-\rnd M(x)}{-\epsilon/(|\opt{x}|\ln(1-\epsilon))}\).  This
gives us our pessimistic estimator: $E[\cost(x_f)+\dist(x_f) \giv x] \le
\pe(x)$ where
\[\pe(x) \doteq \cost(x)+\dist(x)+(k+d)\frac{(1-\epsilon)N-\rnd M(x)}{-\epsilon/\ln(1-\epsilon)}.\]
The algorithm chooses $f$ and $C$ to keep $\pe$ from increasing
(although not necessarily to {\em minimize} $\pe$) at each round.

\begin{guarantee}\label{FFLAG}
  Let $\dtm x$ be the output of the algorithm shown in Fig.~\ref{FFLA}.
  Then $\dtm x$ is a fractional solution to the facility location LP
  and $\cost(\dtm x) + \dist(\dtm x) < (1-\epsilon)^{-1}
  \min_x \dist(x)+\cost(x)$,
  where $x$ ranges over all fractional solutions.
\end{guarantee}
\begin{proof} (Sketch)
  Define $\pe$ as above.
  The algorithm maintains the invariant $\pe(x) \le (k+d)N$.
  A calculation\footnote{%
    This proof is an adaptation of part of the Chernoff-Wald proof
    to this context.  For further details on parts marked with this footnote,
    see that proof.}
  shows that the invariant is initially true by the choice of $N$.
  Clearly if the invariant is true at the end the performance guarantee holds.
  In a given iteration, the increase in $\pe$
  is at most\addtocounter{footnote}{-1}\footnotemark{} $(k+d)\epsilon$ times
  \[\frac{\cost(f)+\sum_{c\in C} \dist(f,c)}{k+d}
  -\frac{\sum_{c\in C} y(c)}{\sum_c y(c)}\]
  where $y(c) = (1-\epsilon)^{x(c)}$.
  If $f$ and $C$ were chosen randomly as in the rounding scheme,
  the expectation of the above quantity would be
  non-positive.\addtocounter{footnote}{-1}\footnotemark{}
  Thus, to keep it non-positive, it suffices to choose
  $f$ and $c$ to maximize
  \[\frac{\sum_{c\in C} y(c)}{\cost(f)+\sum_{c\in C} \dist(f,c)}\]
  which is what the algorithm does.
  \hfill\qed
\end{proof}

In each iteration, at least one customer $c$ with $x(c) \le N$ has $x(c)$
incremented.  Thus the number of iterations is $O(n\ln(n)/\epsilon^2)$.
Each iteration can be implemented in linear times $O(\ln(n))$ time.
Thus,
\begin{corollary}
  Let $0<\epsilon<1$.   Fractional facility location
  has a $[(1+\epsilon)(d+k)]$-approximation algorithm that runs in
  $O(n\ln(n)/\epsilon^2)$ iterations, each requiring time linear in the input
  size times $\ln n$.
\end{corollary}

\section{Further directions}
Is there a {\em greedy} $[d + H_{\Delta} k]$-approximation algorithm for
facility location?  A $[(1+\epsilon)d;(1+\ln(n+n/\epsilon))k]$-approximation
algorithm for {\em weighted} $k$-medians?  A $[d,(1+\epsilon) k]$- or
$[(1+\epsilon)d,k]$-approximation algorithm for fractional $k$-medians?  A
Lagrangian-relaxation algorithm for fractional {\em weighted} $k$-medians?

The running times of all of the algorithms here can probably be improved using
techniques similar to the one that Fleischer applied to improve Garg and
Konemann's multicommodity flow algorithm \cite{FleischerU99}, or (depending on the application)
using standard data structures.

In practice, changing the objective function of the LP relaxation of the IP
to better reflect the performance guarantee might be worthwhile.
For example, if one is going to randomly round a fractional solution $x$
to the facility-location LP, it might be better to minimize
$\dist(x) + \sum_f \cost(f) x(f) H_{\Delta_{f}}$
rather than $\dist(x)+\cost(x)$.
This gives a performance guarantee that is provably as good,
and may allow the LP to compensate for the fact that the difficulty
of approximating the various components of the cost varies.

\section*{Acknowledgements}
Thanks to
Sanjeev Arora,
Lisa Fleischer,
Naveen Garg,
and
David Shmoys
for helpful comments.


\appendix \small

\section{$K$-medians via $k$-set cover via PST. }
For completeness, we discuss a relation between fractional $k$-medians and the
mixed/packing covering framework of Plotkin, Shmoys, and Tardos (PST)
\cite{PlotkinST94}.  First we consider the $k$-set cover problem --- a variant
of weighted set cover in which each set $s$ is given a ``distance''
$\dist(s)\in\Rp$, and the goal is to choose a cover (a collection of sets
containing all elements) of size at most $k$, minimizing the total distance.

We formulate the {\em decision} problem (given $d$, is there a set cover of
size at most $k$ and distance at most $d$?)  as a mixed packing/covering
problem.  Let \(P \doteq \{ x : \sum_s x_s \le K \}\).  For $x\in P$ define
$x_e \doteq \sum_{s\ni e} x_s$ and $\dist(x) \doteq \sum_s x_s\dist(s)$.  Then
the fractional $K$-set cover problem is the packing/covering problem \(\exists?
x \in P : (\forall e) x_e \ge 1; \dist(x) \le d.\)

We can solve this using the PST algorithm as follows.
The input to that algorithm is $(\dist(), k, d, \epsilon)$.
We can use it to compute an approximate solution
$x$ such that $(\forall e) x_e \ge 1-\epsilon$
and $\dist(x) \le (1+\epsilon) d$
(provided the original problem is feasible).
We scale $x$, multiplying it by $1+O(\epsilon)$,
to get the final output.

With care, we can show that to implement the PST algorithm,
it suffices to have a subroutine that, given a vector
$\alpha$, returns $x\in P$
minimizing $\dist(x)-\sum_e \alpha_e x_e$.
An optimal $x$ can be found by enumerating the sets
and choosing the set $s$ that minimizes
\(\dist(s) - \sum_{e\in s} \alpha_e.\)

The running time of the PST algorithm is dominated
by the time spent in this subroutine.  The subroutine is called
$O(\rho \ln(m)/\epsilon^2)$ times,
where $m$ is the number of elements and  $\rho$ is the {\em width} of the problem instance,
which in this case is
\(k \,\max_{s} \{\dist(s)/d, 1 \}.\)
Thus, 
\begin{corollary}
  The fractional $k$-set cover decision problem
  reduces to a mixed packing/covering problem of width
  \(\rho = k \,\max_{s} \{\dist(s)/d, 1 \}\).
  If a problem instance is feasible,
  the algorithm of \cite{PlotkinST94} yields
  a fractional solution $x$ with $|x| \le (1+\epsilon)k$
  and $\dist(x) \le (1+\epsilon) d$ in time linear in the input size times
  $O(\rho \ln(m)/\epsilon^2)$. 
\end{corollary}
In many cases, we can assume without loss of generality
that $\max_s \dist(s) \le d$, in which case the width is $k$.
Except for the fact that this is a decision procedure,
this is comparable to Corollary~\ref{fkm}.
(Although that bound requires no assumption about $d$.)

Next we sketch how weighted $k$-medians reduces to $k$-set cover.
We adapt Hochbaum's facility-location-to-set-cover reduction.
Fix a weighted $k$-medians instance with $n$ facilities and $m$ customers $\cal C$.
Construct an (exponentially large) family of sets as follows.
For each facility $f$ and subset $C$ of customers, define a set $S_{fC} = C$,
with $\dist(S_{fC}) \doteq \sum_{c\in C} \dist(f,c)$ and $\cost(S_{fC}) \doteq \cost(f)$.
Then each $k$-medians solution
corresponds to a $k$-set cover, and vice versa.
The bijection preserves $\dist$,
and extends naturally to the fractional problems as well.

Even though the resulting fractional $k$-set is exponentially large,
we can still solve it efficiently using PST provided we have a subroutine that, given a vector $\alpha$,
efficiently finds a facility $f$ and set of customers $C$ minimizing $\sum_{c\in C} \, \dist(f,c)-\alpha_c$.
This $C$ and $f$ can in fact be found by choosing the facility $f$ minimizing
\(\sum_{c\in C_f} \, \dist(f,c)-\alpha_c\),
where \(C_f = \{ c : \dist(s) < \alpha_c.\}\).
Thus, we have 
\begin{corollary}
  The fractional weighted $k$-medians decision problem
  reduces to a mixed packing/covering problem of width
  \(\rho = k \,\max_{f} \{\sum_c \dist(f,c)/d, 1 \}\).
  If a problem instance is feasible,
  the algorithm of \cite{PlotkinST94} yields
  a fractional solution $x$ with $|x| \le (1+\epsilon)k$
  and $\dist(x) \le (1+\epsilon) d$ in time linear in the input size times
  $O(\rho \ln(m)/\epsilon^2)$. 
\end{corollary}
If each $\dist(f,c) \le d$,
then $\rho \le km$, where $m$ is the number of customers.
This bound on the running time is a factor of $m$
worse than the bound in Corollary~\ref{fkm}
(though a reduction yielding smaller width may be possible).

\end{document}